\documentstyle[sprocl]{article}

\def\Journal#1#2#3#4{{#1} {\bf #2}, #3 (#4)}
\begin{document}

\title{ LONG RANGE INTERACTION CORRECTIONS ON THE QUANTUM VIBRONIC SOLITON}

\author{ D. GRECU, A.S. C\^ARSTEA, ANCA VI\c SINESCU}

\address{Institute of Physics and Nuclear Engineering -
"Horia Hulubei"\\
Bucharest, ROMANIA\\
E-mail:dgrecu@theor1.ifa.ro}

\maketitle\abstracts{ Self-localized modes in a quantum vibronic 
system, with long
range interaction of Kac-Baker type and interacting nonlinearly
with an acoustical phonon bath, is studied. One works in the
coherent state approximation. Following a procedure of Sarker
and Krumhansl the problem is reduced to a nearest neighbours
one. In the continuum limit the localized state satisfy a mKdV
equation. An approximate expression for its frequency is found.}
\noindent 
Since Davydov's pioneering paper on solitons in
quasi-one-dimensional molecular systems\cite{dav1} the subject
was investigated intensively by many 
authors.\cite{chr} One reason is their possible 
role in storage and energy transport in systems of biological 
interest. Of special interest for the present 
paper is the model put forward by Takeno in the middle of
eighties.\cite{tak1}$^{,}$\cite{tak2} It consists of a nonlinear
vibron system coupled nonlinearly with an harmonic bath
of acoustical phonons. The hamiltonian is 
$\hat H=\hat H_{vib}+\hat H_{ac}+\hat H_{int}$
where
\begin{equation}\label{ham}
\hat H_{vib}=\sum_{n}\left({1\over 2\mu} \hat p^{2}_{n} + v(\hat
q_{n})\right) - {1\over 2} \sum_{m\ne n} J_{mn} \hat q_{n}
\hat q_{m}
\end{equation}
with $v(\hat q) = {1\over 2} \mu \Omega^{2}_{0}\hat q^{2} + \mu
{B\over 4}\hat q^{4}$
and the last term in (\ref{ham}) is describing the long range interaction
(LRI) existing between the vibrons. This is usually taken
 of dipole-dipole type, and although of long range nature, 
is restricted only between nearest neighbours. It is our aim 
to consider a mathematical tractable model of long range interaction, 
 the so-called Kac-Baker model,\cite{kb} for which the $J_{mn}$ is 
exponentially decaying,
$J_{mn}=J{1-r\over 2r} e^{-\gamma\vert m-n\vert},\quad r=e^{-\gamma}. $
The advantage of this model
is that it covers continuously the whole spectrum from nearest neighbour case
$(r\to 0)$ to the Van der Waals limit $(r\to 1)$. Our aim is to investigate 
this dependence on $r$ of the existing localized excitations of (\ref{ham}).

The acoustical phonons are described by an harmonic hamiltonian,
\begin{equation}\label{hfon}
\hat H_{ac}=\sum_{n}\left({1\over 2M}\hat P^{2}_{n}+{1\over 2}
K(\hat u_{n+1} - \hat u_{n})^{2}\right),
\end{equation}
and the interaction between vibrons and phonons by
$$\hat H_{int}=\sum_{n}\lambda V(\hat q_{n})(\hat u_{n+1}-\hat u_{n-1})$$
where we consider the simplest form for $V(\hat q_{n})$, namely 
$V(\hat q_{n})={1\over 2}\hat q^{2}_{n}.$

As is generally believed by the authors working in this field, robust
localized excitations of solitonic type in these systems are well described
in the coherent state approximation. It was used by Takeno \cite{tak2}
and together with a variational principle the quantum problem was reduced
to a classical one. Due to the simple form of the now classical vibron-phonon
interaction the last variables can be eliminated by a Laplace transform.
It results a complicated generalized Langevin equation. A full discussion of
this equation was not yet done. As a first step all the memory
and transient-effect terms are neglected. The resulting equation is:

\begin{equation}\label{sis}
\mu {d^{2}q_{n}\over dt^{2}} + v_{N}^{\prime}(q_{n}) - \sum_{m \ne n}
J_{mn}q_{m} -F(q_{n}) =0
\end{equation}
where
\begin{equation}\label{par}
F(q_{n})={\lambda^2 \over K}V'_{N}(q_{n})(V_{N}(q_{n+1})+V_{N}(q_{n-1})
+2V_{N}(q_{N})),
\end{equation}
and the other notations are those used by Takeno.\cite{tak2}
Looking for solutions of the form
$q_{n}= \Phi_{n}(k;t) \cos (kan-\omega t)$
and separating the $\cos$ and $\sin$ parts in the rotating wave approximation
we get
\begin{equation}\label{fi}
{d^{2}\Phi_{n} \over dt^{2}} + \tilde\Omega^{2} \Phi_{n}-\frac{3}{4}\gamma_{1}
\Phi^3_{n}-RC_{n}+\alpha_{k}\Phi_{n}(\Phi^{2}_{n+1}+\Phi^{2}_{n-1})
=\omega^2\Phi_{n}
\end{equation}
\begin{equation}\label{fi1}
{d\Phi_{n}\over dt} +{R\over 2\omega}S_{n}+(\beta_{k}/2\omega)\Phi_{n}(
\Phi_{n+1}^2-\Phi_{n-1}^2)=0.
\end{equation}
Here~~~~~
$\tilde\Omega^2=\Omega^2-\frac{3}{2}\frac{\lambda^2}{K}(q^{(0)})^2,~~~
\gamma_{1}=\frac{\lambda^2}{\mu K}-B,~~
\quad \gamma_{2}=
\frac{\lambda^2}{2\mu K}, $ \\
$$\quad R=\frac{J}{\mu}\frac{1-r}{2r},~~~~
\alpha_{k}=\frac{1}{2}\gamma_{2}(1+\frac{1}{2}\cos{2ka}),~~~~
\beta_{k}=\frac{1}{2}\gamma_{2}\sin{2ka}.$$
Also we have introduced the quantities,
\begin{eqnarray}\label{cn}
C_{n} &=& \sum_{m\ne} e^{-\gamma \vert m-n \vert} \cos (ka(m-n))
\Phi_{m} \nonumber\\
S_{n} &=& \sum_{m\ne n} e^{-\gamma \vert m-n \vert} \sin (ka(m-n))
\Phi_{m}\nonumber \\
\tilde C_{n}& =& \sum_{m\ne m} e^{-\gamma \vert m-n \vert} sgn(m-n)
\cos (ka(m-n)) \Phi_{m} \nonumber \\
\tilde S_{n} &=& \sum_{m\ne n} e^{-\gamma \vert m-n \vert} sgn(m-n)
\sin (ka(m-n)) \Phi_{m}.
\end{eqnarray}
With these quantities following Sarker and Krumhansl 
\cite{sar} it is easy to find "connection relations"
namely,
\begin{eqnarray}\label{rc}
&&C_{n+1}+C_{n-1}= 2r \cos ka \Phi_{n}-(\Phi_{n+1}+\Phi_{n-1}) 
+2r \cosh \gamma \cos ka C_{n}
\nonumber \\
&&~~~~~~~~~~~~~~~~~~+2 \sinh \gamma \sin ka \tilde S_{n} \nonumber \\
&&\tilde C_{n+1}+C_{n-1}=-(\Phi_{n+1}-\Phi_{n-1}) + 2\cosh \gamma
\cos ka \tilde C_{n}+2\sinh \gamma \sin ka S_{n} \nonumber \\
&&S_{n+1}+S_{n-1} = 2\cosh \gamma \cos ka S_{n} - 2\sinh \gamma
\sin ka \tilde C_{n} \nonumber \\
&&\tilde S_{n+1}+\tilde S_{n-1} = 2r \sin ka \Phi_{n} + 2\cosh
\gamma \cos ka \tilde S_{n} - 2 \sinh \gamma \sin ka C_{n} 
\end{eqnarray}
The equations (\ref{fi}), (\ref{fi1}) together with (\ref{rc})
form a complete system containing only the interaction between
nearest neighbours.

Further, we shall work in the continuum approximation $(na\to x)$. 
The equations
(\ref{fi}) and (\ref{fi1}) become $(x\to x/a)$

\begin{equation}\label{fitt}
\Phi_{tt} + \tilde\Omega^{2} \Phi - RC +{3\over
4}(\gamma_{1}+\frac{8\alpha_{k}}{3})\Phi^{3}-2\alpha_{k}(\Phi\Phi_{x}^2
+\Phi^2\Phi_{xx}) = \omega^2 \Phi,
\end{equation}
and 
\begin{equation}\label{fit}
\Phi_{t} + \frac{R}{2\omega}S+\frac{4\beta_{k}}{2\omega}\Phi^2\Phi_{x}= 0
\end{equation}
and also the connection relations (\ref{rc}) are transformed
into their continuum form.

From the continuum limit of the "connection relations" we can
express $S$ as function of $\Phi$ and the  equation (\ref{fit}) becomes
\begin{equation}\label{mmkdv}
\Phi_{t}+v_{k}(\Phi_{x}+\frac{1}{6}\Phi_{xxx})+\frac{4\beta_{k}}{2\omega}
\Phi^2 \Phi_{x}=0,
\end{equation} 
where
$v_{k} = {R\over 2\omega } {\sinh \gamma \sin ka \over (\cosh \gamma
- \cos ka )^{2}}.$
In the moving reference system with velocity $v_{k}$, $y=x-v_{k}t$, 
and introducing the time variable $\tau=v_{k}t/6$ and scaling $\Phi$ as 
$\Psi=A\Phi$ with $A^2=2\omega v_{k}/4\beta_{k}$, 
(\ref{mmkdv}) becomes the standard modified Korteweg de Vries equation
\cite{ab}
\begin{equation}\label{mkdv}
\Psi_{\tau}+\Psi_{yyy}+6\Psi^2 \Psi_{y}=0
\end{equation}

The one soliton solution of (\ref{mkdv}) is given by \cite{ab}:
\begin{equation}\label{solution}
\Psi=\rho\frac{1}{\cosh{\theta}}~~~~~~~
\theta=\rho y-\rho^3 \tau+\theta_{0}
\end{equation}

The equation (\ref{fitt}) can be transformed in a similar way as (\ref{fit}).
Going in the reference system
moving with velocity $v_{k}$ we get a complicated nonlinear 
differential equation in $\Psi$. Of course the solution (\ref{solution})
does not obey it. But we can use it and integrate over the real
$ \theta $
axis and determine $\omega^2$ from the emerged expression. In this very
approximate way we get:

$$\omega^2=\tilde\Omega^2-R\frac{\sinh{\gamma}\sin^2{ka}+
(1-r\cos{ka})(\cosh{\gamma}\cos{ka}-1)}{(\cosh{\gamma}-\cos{ka})^2}-$$
\begin{equation}
-\frac{1}{A^2}(\rho^2\gamma_{1}
+\frac{3}{8}\rho^2\alpha_{k}-\frac{1}{4}\rho^4\alpha_{k}).
\end{equation}
When $r\to 0$ that is in the 
nearest neighbour approximation, 
we obtain
\begin{equation}
\omega^2\to \tilde\Omega^2-\frac{J}{\mu}(\sin^2{ka}+\cos{ka})-
2\cos{ka}\frac{\lambda^2}{KJ}[\rho^2(\gamma_{1}+\frac{3}{8}\alpha_{k})-
\frac{1}{4}\rho^4\alpha_{k}]
\end{equation}
More details will be presented elsewhere.

{\em\section*{Acknowledgments}
This paper was done under the contract
Nr. 70/1998 with the Romanian Academy. Two of the authors (D.G., A.I.V.)
would like to thank Prof. Valerio Tognetti and the Organizing Committee
of the 6-th Int. Conf. on "Path Integrals from peV to TeV" for
giving the opportunity to attend the Conference. Helpful discussions
with Prof. S. Takeno during the Conference time are kindly acknowledged.}

\end{document}